\documentclass[journal]{IEEEtran}

\usepackage{amsmath,amssymb,amsfonts}
\usepackage{algorithmic}
\usepackage{graphicx}
\usepackage{xcolor}
\usepackage{bm}
\usepackage{algorithm}
\usepackage[colorlinks]{hyperref}

\begin{document}

\title{\LARGE Fluid Reconfigurable Intelligent Surface (FRIS) Enabling\\Secure Wireless Communications}

\author{Xusheng Zhu,
            Kai-Kit Wong,~\IEEEmembership{Fellow,~IEEE},
            Boyi Tang,
            Wen Chen, and
            Chan-Byoung Chae,~\IEEEmembership{Fellow,~IEEE}
\vspace{-9mm}

\thanks{(\emph{Corresponding author: Kai-Kit Wong}).}
\thanks{X. Zhu, K. K. Wong, and B. Tang are with the Department of Electronic and Electrical Engineering, University College London, London, United Kingdom. (e-mail: xusheng.zhu@ucl.ac.uk; kai-kit.wong@ucl.ac.uk; boyi.tang.22@ucl.ac.uk). K. K. Wong is also affiliated with the Yonsei Frontier Lab., Yonsei University, Seoul, 03722 South Korea.}
\thanks{W. Chen is  with the Department of Electronic Engineering,
Shanghai Jiao Tong University, Shanghai 200240, China (e-mail: wenchen@sjtu.edu.cn).}
\thanks{C.-B. Chae is with the School of Integrated Technology, Yonsei University, Seoul 03722, South Korea (e-mail: cbchae@yonsei.ac.kr).}
}

\maketitle

\begin{abstract}
The concept of fluid reconfigurable intelligent surface (FRIS) upgrades the conventional reconfigurable intelligent surface (RIS) paradigm by empowering its reflecting elements with positioning reconfigurability. This letter aims to investigate the use of FRIS to enhance physical-layer security in a system, in which a multi-antenna access point (AP) communicates with a legitimate user device in the presence of an eavesdropper. Unlike RIS with fixed-position elements, FRIS can dynamically select an optimal subset of elements from a larger array of candidate locations. We aim to maximize the secrecy rate by jointly optimizing the AP's transmit beamforming, the selection of FRIS activated elements, and their discrete phase shifts. The resulting problem is a challenging mixed-integer nonlinear program (MINLP), which is NP-hard. To address this, we propose an efficient algorithm based on an alternating optimization (AO) framework. Within this framework, the beamforming subproblem is optimally solved in closed form using the generalized eigenvalue method, while the combinatorial subproblem of joint element selection and discrete phase design is handled via the cross-entropy optimization (CEO) method. Simulation results show that the proposed FRIS design significantly outperforms the conventional RIS counterpart and other baselines, demonstrating the substantial security gains by element positioning as the new degree of freedom (DoF).
\end{abstract}


\section{Introduction}
\IEEEPARstart{W}{ireless} systems face growing security threats due to the broadcast nature of the radio medium. Physical-layer security (PLS) has emerged as a promising paradigm to ensure data confidentiality by exploiting the physical characteristics of wireless channels \cite{b1}. Unlike traditional cryptography, PLS provides information-theoretic security without complex key management \cite{Enwereonye-2025}. Recently, reconfigurable intelligent surfaces (RIS) have been recognized as a transformative technology for creating smart and controllable radio environments \cite{b2}. An RIS, composed of low-cost passive reflecting elements capable of inducing controllable phase shifts \cite{zhu2024on,saixu}, can reshape wireless environments to enhance desired signals and suppress interference, hence significantly boosting secrecy performance in wireless communication systems.

Despite their advantages, conventional RIS architectures are limited by practical constraints, most notably finite-resolution phase control of their elements. The consequence is that RIS in practice fails to meet the expectation of the smart environment it was originally imagined, questioning the initial intention of having relatively low-cost RISs to reduce the burden of base stations (BSs) or access points (APs) \cite{Amri-2022}.

To address this, the concept of fluid antenna system (FAS) \cite{wong-fas2021} can be very useful. FAS represents the paradigm of treating antenna as a reconfigurable physical-layer resource to broaden system design and network optimization \cite{New-2025,Wang-2024}. Also, FAS is hardware-agnostic and can appear in many forms depending on specific implementation techniques \cite{Lu-2025}. Relevant to RIS, it is possible to introduce positioning reconfigurability into the elements to compensate for the loss in phase resolution of the elements and obtain more spatial diversity beyond the number of elements. This has led to the concept of fluid RIS (FRIS) \cite{Xiao-2025,Ghadi-2025,b5}, in which each `fluid' element can optimize its position of reflection. Technically speaking, it can also be interpreted as turning on a subset of elements from a larger number of densely populated elements \cite{Xiao-2025}.

FAS can enable PLS and has been studied in \cite{tang2023fluid}. However, secrecy benefits of using FRIS are not well understood. Yet, it is anticipated that FRIS could reshape the spatial relationship between the legitimate and eavesdropping channels, enabling the system to escape unfavorable propagation conditions and enhance secrecy. Motivated by this, this letter aims to study a FRIS-assisted multiple-input single-output (MISO) system that maximizes the secrecy rate through joint optimization of the AP's beamforming, and the FRIS element selection and phase shifts. Our contributions are summarized as follows:
\begin{itemize}
\item We formulate a secrecy rate maximization problem for the FRIS-assisted MISO system, jointly designing beamforming, element selection, and discrete phase shifts.
\item We propose an efficient alternating optimization (AO) algorithm to solve the resulting mixed-integer nonlinear program (MINLP). The beamformer is obtained in closed form via the generalized eigenvalue method, while element selection and phase design are handled using the cross-entropy optimization (CEO) method.
\item Simulation results are provided to verify that the proposed FRIS scheme achieves substantial secrecy rate gains over conventional RIS and other benchmarks, demonstrating the benefits of elements' positioning reconfigurability.
\end{itemize}

\section{System Model}
\begin{figure}[]
\centering
\includegraphics[width=.5\columnwidth]{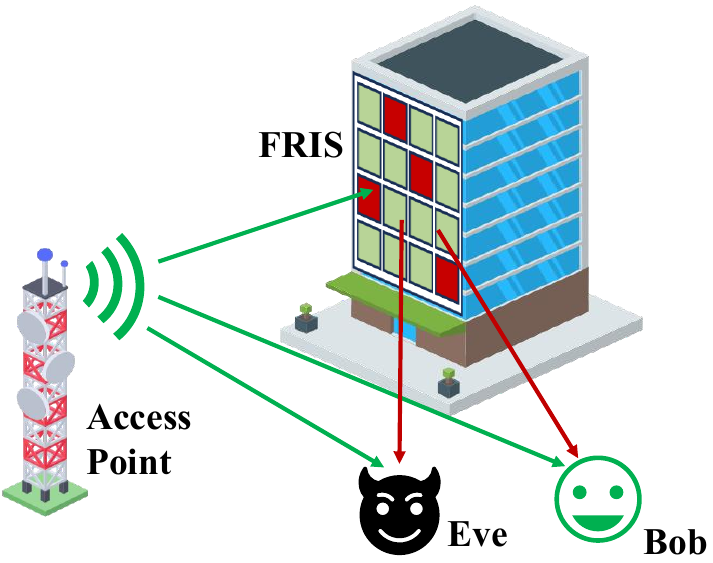}
\caption{A FRIS-assisted secure communication system.}\label{fig_frame}
\vspace{-3mm}
\end{figure}

As illustrated in Fig.~\ref{fig_frame}, we consider a FRIS-assisted secure MISO system where an $M$-antenna AP serves a single-antenna legitimate user terminal (Bob) in the presence of a single-antenna eavesdropper (Eve). All terminals so far are equipped with fixed-position antennas (FPAs). To enhance secrecy rate, a FRIS is deployed. While Fig.~\ref{fig_frame} depicts a planar deployment on a building facade for illustration, this letter focuses on a one-dimensional (1D) linear deployment model. This 1D model is adopted as a standard and tractable approach that captures the fundamental essence of positioning optimization, allowing for the use of established channel correlation models.

\subsection{Channel Model}
\subsubsection{FRIS Geometry and Correlation Model}
The FRIS is modeled as a 1D uniform linear array (ULA) of length $W\lambda$, where $\lambda$ is the carrier wavelength. Along this length, $N$ discrete locations are available, indexed by $n \in \{1, \dots, N\}$, with uniform spacing $d_s = W\lambda / (N-1)$. For dense deployments typical of FRIS, spatial correlation between element channels becomes significant. We adopt the Jakes' model \cite{fasazhu}, widely used for ULAs, to capture this correlation. The correlation coefficient between locations $i$ and $j$ is given as
\begin{equation}
[\mathbf{R}]_{i,j} = J_0\left(\frac{2\pi d_{i,j}}{\lambda}\right) = J_0\left(2\pi |i-j| \frac{d_s}{\lambda}\right),
\end{equation}
where $d_{i,j} = |i-j|d_s$ is the physical distance between the $i$-th and $j$-th locations, and $J_0(\cdot)$ is the zeroth-order Bessel function of the first kind. This results in an $N \times N$ symmetric Toeplitz correlation matrix $\mathbf{R}$, assumed to be common for all FRIS-related links (AP-FRIS, FRIS-Bob, and FRIS-Eve).

\subsubsection{Channel Definitions}
All channels are modeled as quasi-static, incorporating large-scale path loss, spatial correlation introduced by FRIS, and small-scale fading. This quasi-static assumption is valid for scenarios where the channel coherence time is much larger than the transmission block length. We justify our specific modeling choices below. A key scenario motivating this scenario is one where the direct AP-user links are obstructed, thereby necessitating the FRIS-scattered path to establish a secure and reliable connection.

The direct AP-Bob and AP-Eve channels, $\mathbf{h}_{d,B} \in \mathbb{C}^{M \times 1}$ and $\mathbf{h}_{d,E} \in \mathbb{C}^{M \times 1}$, are modeled, respectively, as
\begin{align}
\mathbf{h}_{d,B} &= \sqrt{\beta_{d,B}} \tilde{\mathbf{h}}_{d,B}, \\
\mathbf{h}_{d,E} &= \sqrt{\beta_{d,E}} \tilde{\mathbf{h}}_{d,E},
\end{align}
in which $\beta_{d,B}$ and $\beta_{d,E}$ denote the respective path loss values, and $\tilde{\mathbf{h}}_{d,B}, \tilde{\mathbf{h}}_{d,E} \in \mathbb{C}^{M \times 1}$ are independent and identically distributed (i.i.d.) complex Gaussian vectors, $\mathcal{CN}(\mathbf{0}, \mathbf{I})$.

Considering that the typical deployment location of AP-FRIS is usually at a high position, the AP-FRIS link is likely to contain a stable line-of-sight (LoS) component. In this manner, the $N \times M$ channel matrix $\mathbf{G}$ from the AP to the $N$ FRIS locations is modeled as
\begin{equation}
\mathbf{G} = \sqrt{\beta_{AP}} \left( \sqrt{\frac{K}{K+1}} \mathbf{G}_{\text{LoS}} + \sqrt{\frac{1}{K+1}} \mathbf{R}^{1/2} \mathbf{G}_{\text{NLoS}} \right),
\end{equation}
where $\beta_{AP}$ represents the large-scale path loss, $K$ denotes the Rician K-factor, $\mathbf{G}_{\text{LoS}}$ is the deterministic LoS component, $\mathbf{G}_{\text{NLoS}} \in \mathbb{C}^{N \times M}$ contains i.i.d.~$\mathcal{CN}(0,1)$ entries for the non-LoS component, and $\mathbf{R}^{1/2}$ is the Cholesky decomposition of the correlation matrix $\mathbf{R} = \mathbf{R}^{1/2} (\mathbf{R}^{1/2})^H$.

We assume a rich scattering environment between the FRIS and the ground-level users (Bob and Eve). Thus, the channels from the $N$ FRIS locations to Bob and Eve are modeled as
\begin{align}
\mathbf{h}_{r,B} &= \sqrt{\beta_{B}} \mathbf{R}^{1/2} \tilde{\mathbf{h}}_{r,B}, \\
\mathbf{h}_{r,E} &= \sqrt{\beta_{E}} \mathbf{R}^{1/2} \tilde{\mathbf{h}}_{r,E},
\end{align}
where $\beta_{B}$ and $\beta_{E}$ are the respective path loss values, and $\tilde{\mathbf{h}}_{r,B}, \tilde{\mathbf{h}}_{r,E} \in \mathbb{C}^{N \times 1}$ are i.i.d.~$\mathcal{CN}(\mathbf{0}, \mathbf{I})$ vectors.

\subsection{Signal Model and Secrecy Performance}
The FRIS controller selects a subset of $\hat{N}$ elements from the $N$ available locations to be activated, where $\hat{N} \ll N$. This selection is represented by a binary vector $\mathbf{s} = [s_1, \dots, s_N]^T$, where $s_n=1$ if the $n$-th element is activated and $s_n=0$ otherwise, subject to $\sum_{n=1}^N s_n = \hat{N}$.

The $n$-th activated element applies a discrete phase shift $\theta_n \in \mathcal{F}$, where $\mathcal{F} = \{0, \frac{2\pi}{2^B}, \dots, \frac{2\pi(2^B-1)}{2^B}\}$ is the set of $2^B$ available phases. The FRIS response is captured by a diagonal matrix $\mathbf{\Phi} = {\rm diag}(s_1 e^{j\theta_1}, \dots, s_N e^{j\theta_N})$.

The AP transmits the signal $x$ assuming $\mathbb{E}[|x|^2]=1$ using a beamforming vector $\mathbf{w} \in \mathbb{C}^{M \times 1}$, subject to the power constraint $\|\mathbf{w}\|^2 \le P_{AP}$. Consequently, the signals received at Bob and Eve are given by
\begin{align}
y_B &= (\mathbf{h}_{d,B}^H + \mathbf{h}_{r,B}^H \mathbf{\Phi} \mathbf{G}) \mathbf{w} x + n_B, \\
y_E &= (\mathbf{h}_{d,E}^H + \mathbf{h}_{r,E}^H \mathbf{\Phi} \mathbf{G}) \mathbf{w} x + n_E,
\end{align}
where $n_B, n_E \sim \mathcal{CN}(0, \sigma^2)$ are the additive white Gaussian noise (AWGN) at Bob and Eve. The corresponding signal-to-noise ratios (SNRs) are given by
\begin{align}
\gamma_B &= \frac{|(\mathbf{h}_{d,B}^H + \mathbf{h}_{r,B}^H \mathbf{\Phi} \mathbf{G})\mathbf{w}|^2}{\sigma^2},\label{eq:rB}\\
\gamma_E &= \frac{|(\mathbf{h}_{d,E}^H + \mathbf{h}_{r,E}^H \mathbf{\Phi} \mathbf{G})\mathbf{w}|^2}{\sigma^2}.\label{eq:rE}
\end{align}
As a result, the achievable secrecy rate $R_s$ which is the maximum of the difference between the legitimate rate and the eavesdropper's rate, can be obtained as
\begin{equation}
R_s = [\log_2(1 + \gamma_B) - \log_2(1 + \gamma_E)]^+,
\end{equation}
where $[x]^+ \triangleq \max(x, 0)$. To establish a theoretical performance upper bound, perfect channel state information (CSI) for all links, including those to the eavesdropper, is assumed to be available at the AP for optimization.

\section{Proposed Algorithm}
\subsection{Problem Formulation}
Our objective is to maximize the secrecy rate $R_s$. This is achieved by jointly optimizing the AP's transmit beamforming vector $\mathbf{w}$, the FRIS element selection vector $\mathbf{s}$, and the phase shift vector $\boldsymbol{\theta}$. Leveraging the definitions of $\gamma_B$ and $\gamma_E$ from (\ref{eq:rB}) and (\ref{eq:rE}), and recognizing the monotonicity of the $\log_2(\cdot)$ function, the problem is formulated as
\begin{subequations}
\label{prob:p1}
\begin{align}
(\text{P1}): \max_{\mathbf{w}, \mathbf{s}, \boldsymbol{\theta}} & \quad \frac{1 + \gamma_B}{1 + \gamma_E} \label{eq:objective} \\
\text{s.t.} & \quad \|\mathbf{w}\|^2 \le P_{AP}, \label{eq:power_constraint}\\
& \quad s_n \in \{0,1\},~\forall n \in \{1, \dots, N\}, \label{eq:binary_constraint} \\
& \quad \sum_{n=1}^N s_n = \hat{N}, \label{eq:cardinality_constraint} \\
& \quad \theta_n \in \mathcal{F},~\forall n \in \{1, \dots, N\}, \label{eq:phase_constraint}
\end{align}
\end{subequations}
where $\mathbf{\Phi} = {\rm diag}(s_1 e^{j\theta_1}, \dots, s_N e^{j\theta_N})$ is the FRIS response matrix. Problem (P1) is a challenging MINLP. The difficulties arise from the non-convex nature of the fractional objective function \eqref{eq:objective}, the intricate coupling of the continuous variable $\mathbf{w}$ with the discrete variables $(\mathbf{s}, \boldsymbol{\theta})$, and the combinatorial search space of $\binom{N}{\hat{N}}(2^B)^{\hat{N}}$ for the FRIS configuration.

\subsection{Proposed AO Framework}
To address (P1), we propose an efficient algorithm based on the AO framework. The idea is to decompose the original problem into two subproblems, which are then solved iteratively until convergence. We alternate between optimizing the transmit beamformer $\mathbf{w}$ and the FRIS configuration $(\mathbf{s}, \boldsymbol{\theta})$.

\subsubsection{Transmit Beamforming Optimization}
For a fixed FRIS configuration $(\mathbf{s}, \bm{\theta})$, the phase shift matrix $\mathbf{\Phi}$ is determined. The effective end-to-end channels to Bob and Eve are respectively defined as
\begin{align}
\mathbf{h}_B^H &= \mathbf{h}_{d,B}^H + \mathbf{h}_{r,B}^H \mathbf{\Phi} \mathbf{G}, \\
\mathbf{h}_E^H &= \mathbf{h}_{d,E}^H + \mathbf{h}_{r,E}^H \mathbf{\Phi} \mathbf{G}.
\end{align}
The subproblem for optimizing $\mathbf{w}$ is then given by
\begin{subequations}
\label{eq:P2}
\begin{align}
(\text{P2}): \max_{\mathbf{w}} \quad & \frac{\sigma^2 + |\mathbf{h}_B^H \mathbf{w}|^2}{\sigma^2 + |\mathbf{h}_E^H \mathbf{w}|^2} = \frac{\sigma^2 + \mathbf{w}^H \mathbf{h}_B \mathbf{h}_B^H \mathbf{w}}{\sigma^2 + \mathbf{w}^H \mathbf{h}_E \mathbf{h}_E^H \mathbf{w}} \\
\text{s.t.} \quad & \|\mathbf{w}\|^2 \le P_{AP}.
\end{align}
\end{subequations}
This is a standard fractional programming problem, and it is well known that the optimal solution must satisfy the power constraint with equality, i.e., $\|\mathbf{w}\|^2 = P_{AP}$. By letting $\mathbf{w} = \sqrt{P_{AP}}\bar{\mathbf{w}}$ where $\|\bar{\mathbf{w}}\|=1$, and defining $\mathbf{R}_B = \mathbf{h}_B \mathbf{h}_B^H$ and $\mathbf{R}_E = \mathbf{h}_E \mathbf{h}_E^H$, the objective function becomes
\begin{equation}
\frac{\sigma^2 + P_{AP} \bar{\mathbf{w}}^H \mathbf{R}_B \bar{\mathbf{w}}}{\sigma^2 + P_{AP} \bar{\mathbf{w}}^H \mathbf{R}_E \bar{\mathbf{w}}} = \frac{\bar{\mathbf{w}}^H (P_{AP}\mathbf{R}_B + \sigma^2\mathbf{I}) \bar{\mathbf{w}}}{\bar{\mathbf{w}}^H (P_{AP}\mathbf{R}_E + \sigma^2\mathbf{I}) \bar{\mathbf{w}}}.
\end{equation}
This is in the form of a generalized Rayleigh quotient. The optimal unit-norm beamformer $\bar{\mathbf{w}}^*$ that maximizes this ratio is the eigenvector corresponding to the maximum generalized eigenvalue of the matrix pair $( (P_{AP}\mathbf{R}_B + \sigma^2\mathbf{I}), (P_{AP}\mathbf{R}_E + \sigma^2\mathbf{I}) )$. Therefore, the optimal transmit beamformer for this subproblem can be found in closed form.

\subsubsection{FRIS Configuration Optimization}
With a fixed beamformer $\mathbf{w}$, the subproblem for optimizing the FRIS configuration $(\mathbf{s}, \boldsymbol{\theta})$ can be formulated as
\begin{subequations}
\label{prob:p3}
\begin{align}
(\text{P3}): \max_{\mathbf{s}, \boldsymbol{\theta}} & \quad \frac{1 + \gamma_B}{1 + \gamma_E} \\
\text{s.t.} & \quad \eqref{eq:binary_constraint}, \eqref{eq:cardinality_constraint}, \eqref{eq:phase_constraint}.
\end{align}
\end{subequations}
This remains an NP-hard problem. We propose to solve it by using the CEO method, a powerful meta-heuristic algorithm well suited for such discrete optimization tasks.

\subsection{FRIS Configuration via CEO}
The subproblem (P3) is a MINLP with a search space of size $\binom{N}{\hat{N}} 2^{B{\hat{N}}}$, making an exhaustive search computationally prohibitive. To tackle this, we employ the CEO method \cite{b6}, which iteratively learns an optimal probability distribution over the solution space to generate high-quality candidate solutions.

\subsubsection{Sampling Distribution Model}
Let a FRIS configuration be $\Gamma = (\mathcal{I}, \boldsymbol{\psi})$, with $\mathcal{I}$ as the index set of $\hat{N}$ activated elements and $\boldsymbol{\psi}$ as their phase shifts. We define a parameterized probability distribution $f(\Gamma; \boldsymbol{\Xi})$ to generate configurations. The CEO algorithm iteratively updates $f(\Gamma; \boldsymbol{\Xi})$ to approximate an ideal distribution $g^*(\Gamma)$ concentrated on the optimal solution $\Gamma^*$ by minimizing the Kullback-Leibler (KL) divergence as
\begin{equation}
\min_{\boldsymbol{\Xi}} D_{KL}(g^* \| f) = \min_{\boldsymbol{\Xi}} \mathbb{E}_{g^*} \left[ \ln \frac{g^*(\Gamma)}{f(\Gamma; \boldsymbol{\Xi})} \right].
\end{equation}
In practice, we simplify the parameter set $\boldsymbol{\Xi}$ to a probability mass function (PMF) vector $\mathbf{p} = [p_1, \dots, p_N]^T$ governing element selection, while choosing phases randomly from $\mathcal{F}$.

\subsubsection{Iterative Solution Generation and Distribution Update}
The CEO algorithm proceeds via the following iterative steps:

{\bf Sample Generation:} At iteration $t$, draw $K$ random samples $\{\Gamma_k\}_{k=1}^K$ from the distribution defined by PMF $\mathbf{p}^{(t-1)}$. For each sample $\Gamma_k$: (i) The set of active element indices, $\mathcal{I}_k$, is generated by drawing $\hat{N}$ indices without replacement from $\{1, \dots, N\}$ according to the learnable PMF $\mathbf{p}^{(t-1)}$; and (ii) The phase shifts for these selected elements are chosen uniformly at random from the discrete set $\mathcal{F}$.

{\bf Evaluation and Selection:} Evaluate the objective function $O(\Gamma_k)$ for each sample. After sorting the samples by performance, the top $K_e = \lceil \rho K \rceil$ samples are selected to form the elite set $\mathcal{E}$, where $\rho$ is the elite ratio.

{\bf Parameter Update:} The PMF $\mathbf{p}$ is updated via a maximum likelihood estimation based on the elite set $\mathcal{E}$ as
\begin{subequations}
\begin{align}
    \max_{\mathbf{p}} &\quad \frac{1}{K_e} \sum\nolimits_{\Gamma_k \in \mathcal{E}} \ln f(\Gamma_k; \mathbf{p}) \\
    \text{s.t.} & \quad \sum\nolimits_{n=1}^N p_n = 1, \\
    & \quad p_n \ge 0,~\forall n.
\end{align}
\end{subequations}
The Lagrangian for this problem becomes
\begin{equation}
    \mathcal{L}(\mathbf{p}, \lambda) = \sum_{\Gamma_k \in \mathcal{E}} \sum_{n \in \mathcal{I}_k} \ln p_n + \lambda \left(1 - \sum\nolimits_{n=1}^N p_n\right),
\end{equation}
where $\lambda$ is the Lagrange multiplier. Solving this yields the update rule for the intermediate PMF $\hat{\mathbf{p}}^{(t)}$ as
\begin{equation}
\hat{p}_n^{(t)} = \frac{\sum_{\Gamma_k \in \mathcal{E}} \mathbb{I}(n \in \mathcal{I}_k)}{\sum_{j=1}^{N} \sum_{\Gamma_k \in \mathcal{E}} \mathbb{I}(j \in \mathcal{I}_k)} = \frac{1}{K_e \hat{N}} \sum_{\Gamma_k \in \mathcal{E}} \mathbb{I}(n \in \mathcal{I}_k),
\end{equation}
where $\mathbb{I}(\cdot)$ is the indicator function. The optimal update is the empirical frequency of each element's appearance in the elite set. A smoothing step is then applied to prevent premature convergence using
\begin{equation}
\mathbf{p}^{(t)} = (1-\alpha) \mathbf{p}^{(t-1)} + \alpha \hat{\mathbf{p}}^{(t)},
\end{equation}
in which $\alpha \in (0, 1]$ represents a smoothing parameter. This process repeats until convergence or a maximum number of iterations is reached.

\subsection{Overall Algorithm and Complexity Analysis}
The proposed AO-CEO algorithm is summarized by initializing the FRIS configuration $(\mathbf{s}^{(0)}, \boldsymbol{\theta}^{(0)})$, and then iteratively executing the two subproblems. In each iteration $i$, we first solve (P2) to obtain $\mathbf{w}^{(i)}$ using the fixed $(\mathbf{s}^{(i-1)}, \boldsymbol{\theta}^{(i-1)})$. Then we solve (P3) using the CEO algorithm with the fixed $\mathbf{w}^{(i)}$ to obtain $(\mathbf{s}^{(i)}, \boldsymbol{\theta}^{(i)})$. The objective value of (P1) is non-decreasing in each iteration, guaranteeing the convergence of the algorithm to at least a local optimum.

\begin{algorithm}[t]
\caption{AO-CEO for Secrecy Rate Maximization}
{\small
\begin{algorithmic}[1]
\STATE \textbf{Initialize}: FRIS configuration $(\mathbf{s}^{(0)}, \bm{\theta}^{(0)})$, AO iteration index $i=0$.
\REPEAT
\STATE $i \leftarrow i+1$.
\STATE \textbf{Step 1: Transmit Beamforming Optimization}
\STATE Given $(\mathbf{s}^{(i-1)}, \bm{\theta}^{(i-1)})$, solve (P2) to obtain $\mathbf{w}^{(i)}$ via the generalized eigenvalue method.
\STATE \textbf{Step 2: FRIS Configuration Optimization}
\STATE Given $\mathbf{w}^{(i)}$, solve (P3) using the CEO algorithm to obtain $(\mathbf{s}^{(i)}, \bm{\theta}^{(i)})$.
\UNTIL{the objective value converges or $i \ge I_{\max}$.}
\STATE \textbf{Output}: $\mathbf{w}^* = \mathbf{w}^{(i)}$, $\mathbf{s}^* = \mathbf{s}^{(i)}$, $\bm{\theta}^* = \bm{\theta}^{(i)}$.
\end{algorithmic}
\label{alg1}}
\end{algorithm}

The computational complexity per AO iteration is dominated by solving the two subproblems. The generalized eigenvalue decomposition in Step 1 has a complexity of $\mathcal{O}(M^3)$. The CEO algorithm in Step 2 requires evaluating $K$ samples in each of its $I_{\text{CEO}}$ iterations. Each evaluation, which involves calculating the objective of (P3), has a complexity of $\mathcal{O}(M\hat{N})$. Thus, the complexity of Step 2 is $\mathcal{O}(I_{\text{CEO}} K M \hat{N})$. The total complexity for $I_{\text{AO}}$ iterations is $\mathcal{O}(I_{\text{AO}}(M^3 + I_{\text{CEO}} K M \hat{N}))$.

\begin{figure}[!t]
\centering
\includegraphics[width=2.5in]{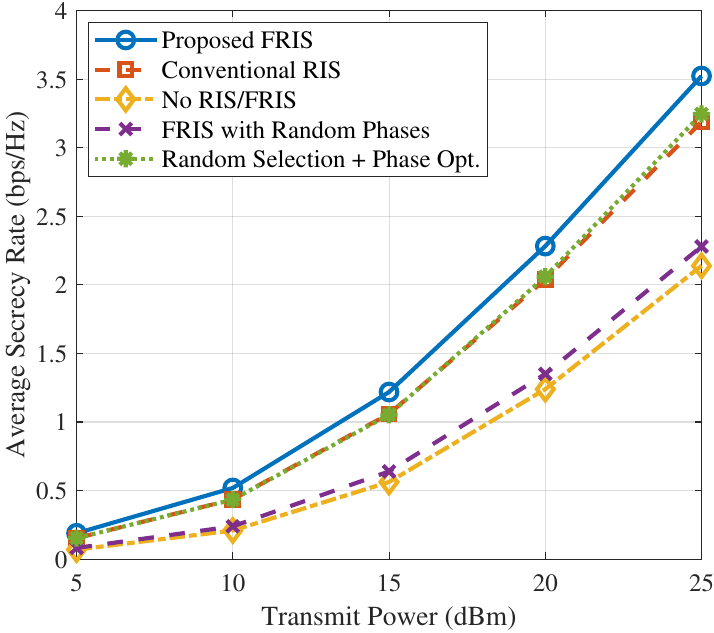}
\caption{Average secrecy rate vs. AP transmit power ($\hat{N}=16$).}\label{fig_vs_power}
\vspace{-3mm}
\end{figure}

\begin{figure}[!t]
\centering
\includegraphics[width=2.5in]{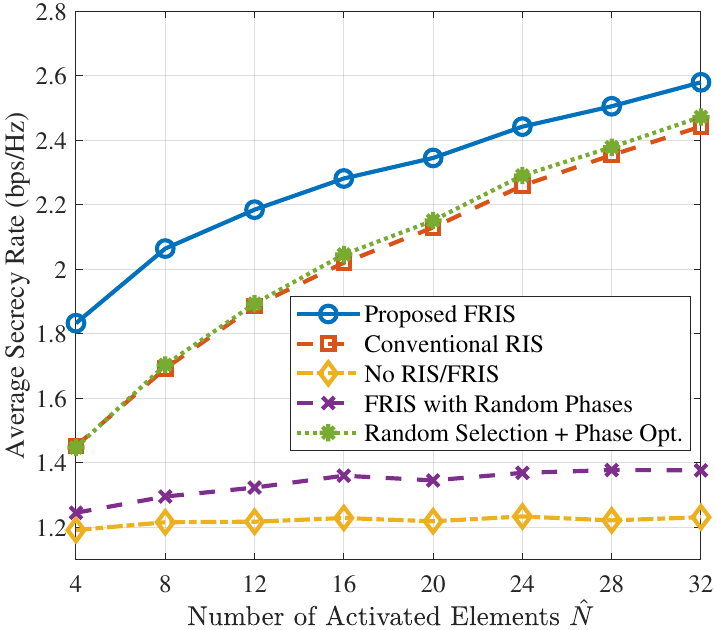}
\caption{Average secrecy rates vs. number of activated elements.}\label{fig_vs_n_hat}
\vspace{-3mm}
\end{figure}

\section{Simulation Results}
Here, we evaluate the proposed design via Monte Carlo simulations. Unless specified otherwise, the system parameters are set as follows. The AP ($M=4$), Bob, and Eve are located at $(0, 0, 10)$ m, $(50, 0, 1.5)$ m, and $(55, 5, 1.5)$ m, respectively. The FRIS, centered at $(45, 10, 5)$ m, has $N=100$ available locations, from which $\hat{N}=16$ elements with $B=3$ quantization bits \cite{zhu2024on} are activated. Additionally, the path loss model is $PL(d) = -30 - 10\alpha \log_{10}(d)$ dB, with exponents $\alpha=2.2$ (AP-FRIS) and $\alpha=2.8$ (all other links). The AP-FRIS link follows Rician fading ($K=5$ dB), while other links are Rayleigh. Crucially, we emulate a challenging NLoS scenario by assuming a $25$ dB additional blockage loss on the direct AP-user paths. The noise power is $\sigma^{2}=-80$ dBm, and the AP transmit power is $P_{AP}= 20$ dBm (unless varied). For the CEO algorithm, we set the sample size dynamically as $K=5N$, with $\rho=0.1$ and $\alpha=0.7$. All results are averaged over $1,000$ independent channel realizations.

Fig.~\ref{fig_vs_power} depicts the average secrecy rate as a function of the AP's transmit power, $P_{AP}$, with $N=100$ and $\hat{N}=16$. As anticipated, the secrecy rates of all schemes exhibit a monotonic increase with $P_{AP}$. The proposed FRIS design consistently achieves the highest secrecy rate across the entire power range, demonstrating a significant performance advantage over all benchmarks. Notably, it substantially outperforms the ``Random Selection + Phase Opt." scheme, which in turn outperforms the ``Conventional RIS". This double gap underscores the twofold benefit: 1) optimizing element positions from a large area is superior to random selection, and 2) even random selection from a large area is superior to a fixed central deployment. The ``FRIS with Random Phases" scheme performs the lowest among the assisted schemes, highlighting that coherent phase optimization is critical for security.

In Fig.~\ref{fig_vs_n_hat}, we study the impact of the number of activated elements, $\hat{N}$, while keeping $N=100$ and $P_{AP}=20$ dBm. The secrecy rates for all schemes incorporating a reflecting surface improve as $\hat{N}$ increases, attributable to the enhanced beamforming gain. Critically, the performance gap between the proposed FRIS and all other benchmarks, particularly the ``Conventional RIS" and ``Random Selection + Phase Opt.", widens progressively with increasing $\hat{N}$. This trend strongly indicates that our joint optimization strategy effectively leverages the larger number of activated elements to better exploit the spatial degrees of freedom, thereby establishing a more dominant channel advantage for Bob relative to Eve.

\begin{figure}[!t]
\centering
\includegraphics[width=2.6in]{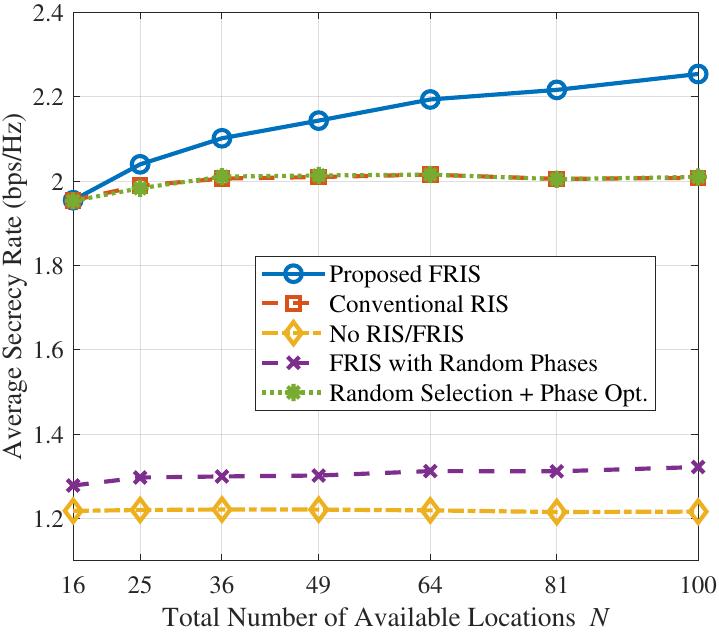}
\caption{Average secrecy rate vs. total available locations $N$ ($\hat{N}=16$).}\label{fig_vs_n_total}
\vspace{-3mm}
\end{figure}

Fig.~\ref{fig_vs_n_total} showcases the fundamental benefit of the FRIS concept by evaluating the secrecy rate as a function of the total number of available locations, $N$, with $\hat{N}=16$. As expected, the performance of the ``No RIS/FRIS" schemes remains constant, being independent of $N$. In striking contrast, the secrecy rate achieved by the proposed FRIS scheme demonstrates a significant monotonic increase with $N$. This result compellingly illustrates that expanding the pool of candidate locations, i.e., a more fluid surface, enhances the spatial selection diversity, enabling the CEO algorithm to identify a more advantageous subset of element positions that maximizes the secrecy performance. Interestingly, at the edge case $N=\hat{N}=16$, the FRIS concept reduces to a conventional one, and the performance of all phase-optimized schemes converges, which is consistent with theory.

\begin{figure}[!t]
\centering
\includegraphics[width=2.6in]{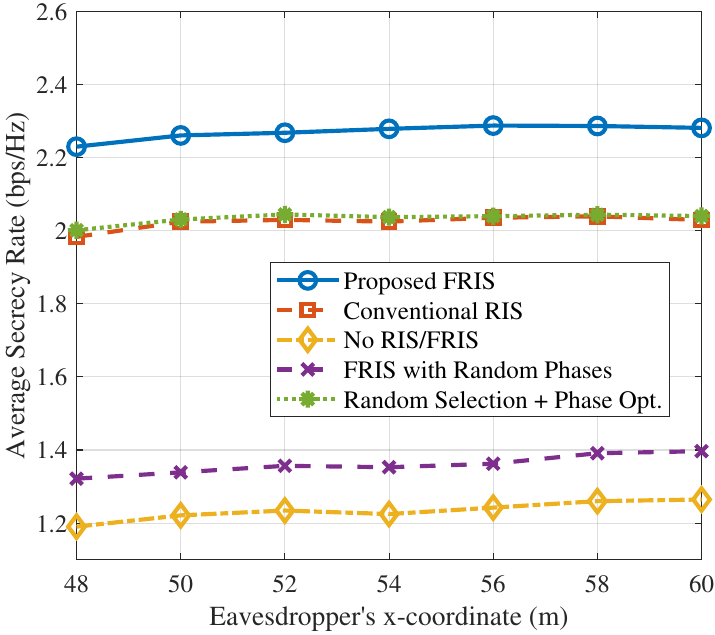}
\caption{Average secrecy rate vs. eavesdropper's location ($\hat{N}=16$).}\label{fig_vs_eve_pos}
\vspace{-3mm}
\end{figure}

Finally, Fig.~\ref{fig_vs_eve_pos} assesses the robustness of the proposed system against variations in the eavesdropper's location. We vary Eve's x-coordinate from $48$ m to $60$ m, altering its proximity to Bob (at $x=50$ m). As Eve moves closer to Bob, the spatial separation decreases, and the secrecy rates of all schemes degrade. Nevertheless, the proposed FRIS (AO-CEO) scheme consistently maintains the highest secrecy rate across all tested locations, exhibiting superior resilience. This demonstrates the effectiveness of jointly optimizing element positions and phases in adapting to spatially challenging scenarios and actively suppressing information leakage.

\section{Conclusion}
This letter proposed a novel approach to enhance PLS using FRIS. We formulated the problem of maximizing the secrecy rate by jointly optimizing the AP's transmit beamforming, and the FRIS's element selection and discrete phase shifts. We devised an efficient AO algorithm, where the FRIS configuration is optimized via the CEO method. Comprehensive simulation results showed the superiority of our proposed FRIS system, which achieves significant secrecy rate improvements over conventional RIS-based systems. This work reveals that the positioning degrees of freedom offered by FRIS are a powerful new resource for securing wireless communications. Future work will focus on robust design under imperfect CSI and the development of low-overhead channel acquisition schemes.

\end{document}